\documentstyle[preprint,aps,eqsecnum,epsfig]{revtex}
\tighten
\begin{document}
\preprint{LPTHE-98-14; DEMIRM-98} 
\renewcommand{\theequation}{\thesection.\arabic{equation}}
\title{\bf Quantum String Dynamics in the conformal invariant SL(2,R) WZWN
Background: Anti-de Sitter Space with Torsion}
\author{\bf H. J. de Vega$^{(a)}$, A. L. Larsen$^{(b)}$ and
N. S\'{a}nchez$^{(c)}$} 
\address
{ (a) LPTHE, Laboratoire Associ\'{e} au CNRS UA280,
Universit\'{e} Pierre et Marie Curie (Paris VI) et Denis Diderot  (Paris VII),
Tour 16, 1er \'{e}tage, 4, Place Jussieu,
75252 Paris, cedex 05, France.\\
(b) Department of Physics, University of Odense,
Campusvej 55, 5230 Odense M, Denmark.\\
(c)Observatoire de Paris, Demirm\footnote{Laboratoire Associ\'e au CNRS
UA 336, Observatoire de Paris et \'Ecole Normale Sup\'erieure. } , 61, Avenue
de l'Observatoire, 75014 Paris, France.}
\date{March 1998}
\maketitle
\begin{abstract}
We consider classical and quantum strings in the conformally invariant
background corresponding to the SL(2,R) WZWN model. This background is
locally anti-de Sitter spacetime with non-vanishing torsion. Conformal
invariance is expressed as the torsion being parallelized.
The precise effect of the conformal
invariance on the dynamics of both circular and generic classical strings is
extracted. In particular, the conformal invariance gives rise to a repulsive 
interaction of the string with the background which precisely cancels the 
dominant attractive term arising from gravity.

We perform both semi-classical and canonical string-quantization,
in order to see the effect of the conformal invariance of the background
on the string mass spectrum. Both approaches yield that the high-mass states 
are governed by $m\sim HN$ ($N\in N_{0},\;\;N$ `large'), where $m$ is the
string mass and $H$ is the Hubble constant. It follows that the level
spacing grows proportionally to $N: \; d(m^2\alpha')/dN\sim N$, while the 
entropy goes like $ S\sim\sqrt{m} $.  Moreover, it follows that there is no 
Hagedorn temperature, so that the partition function is well-defined at any 
positive temperature.

All results are compared with the analogue results in Anti-de Sitter 
spacetime, which is a non conformal invariant background.
Conformal invariance {\it simplifies} the mathematics of the
problem but the physics remains mainly  {\it unchanged}. Differences between 
conformal and non-conformal backgrounds only appear in the intermediate region
of the string spectrum, but these differences are minor. For low and high 
masses, the string mass spectra in conformal and non-conformal backgrounds  
are identical. 

Interestingly enough, conformal invariance fixes the value of the
spacetime curvature to be $ -69/(26 \; \alpha') $.
\end{abstract}
\section{Introduction}
\setcounter{equation}{0}
The systematic investigation of string dynamics in curved spacetimes started in
Ref.\cite{san1}, has revealed new insights and new physical phenomena with
respect to string propagation in flat spacetime (and with respect to quantum 
fields in curved spacetime)\cite{san2}. These results are relevant both for 
fundamental quantum strings and for cosmic strings, which behave in a 
classical way.

Cosmic strings can be considered in arbitrary curved spacetime backgrounds,
while fundamental quantum strings demand a conformally invariant background 
for quantum consistency (conformal invariance is a
necessary although not sufficient condition for consistency). However, most
curved spacetimes that were
historically of physical interest in general relativity and cosmology are
not conformally invariant. On
the other hand,  certain group-manifolds and coset-spaces provide a large
family of new spacetimes that
are conformally invariant, but they are generally not so interesting from a
physical point of view.

The classical and quantum string dynamics and their associated effects
in a wide class of string backgrounds (conformal and non-conformal invariant)
have been widely investigated by the present authors\cite{san1,san2}.

In this paper, we consider classical and quantum strings in the conformally
invariant background corresponding to the $SL(2,R)$-WZWN model. This 
background is locally Anti de Sitter
spacetime with non-vanishing parallelizing torsion. The cosmological
importance of Anti de
Sitter is somewhat less than that of (say) de Sitter, but it is in any case
an example of a
Robertson-Walker spacetime. Moreover, after a suitable point
identification, the background
corresponds to the $2+1$ Black Hole Anti de Sitter spacetime \cite{ban},
which is a toy-model
for investigations of black hole phenomena in higher dimensions.
Thus, our interest in the $SL(2,R)$-WZWN background is due to  a compromise of
conformal invariance, physical interest and simplicity.

Many mathematical aspects of the $SL(2,R)$-WZWN model have been discussed in
the literature (see for
instance Refs.\cite{bal,pet,nem,hwa}), but we find that the physical 
aspects have not really been
extracted so far. The purpose of this paper is to investigate directly the
effect of the conformal
invariance on the string dynamics, both classically and quantum
mechanically. The conformal invariance
is expressed via a parallelizing torsion. Thus we consider the string
equations of motion in a
background consisting of the standard Anti de Sitter metric plus an
anti-symmetric tensor representing
the parallelizing torsion. By considering special as well as generic
solutions to these equations,
and by comparing with the analogue results in the absence of torsion, we
extract the {\it precise}
effect of the conformal invariance on the dynamics of classical strings.
Similarly, after quantization,
we extract the effect of the conformal invariance on the quantum phenomena,
especially on phenomena related
to the quantum mass spectrum.

In the cases of AdS and BH-AdS, the torsion corresponding to conformal 
invariance provides a repulsive term, which in the string dynamics precisely 
cancels the dominant attractive term arising from  gravity.

As a general effect, we find that conformal 
invariance simplifies the mathematics of the problem; however, the physics is
more or less {\it unchanged}. In fact, in the two limits
$n<<(H^2\alpha')^{-1}$ and $ n>>(H^2\alpha')^{-1}$, of the string mass 
spectrum the results obtained here are in {\it exact} agreement with the 
results obtained without torsion \cite{vega2}. For small $n$ and 
large $n$, the spectrum is not affected by the conformal invariance, while 
there are some minor changes in the intermediate region.

The frequencies of string oscillators are shifted away from integers $n$:
\begin{equation}\label{corref}
\omega_n=|n\pm mH\alpha'|,
\end{equation}
while  in 2+1 AdS without torsion, the frequencies 
 turned out to be \cite{all2}:
\begin{equation}\label{sintors}
\omega_n=\sqrt{n^2 + m^2H^2\alpha'^2},\;\;\;\;\;\;\;\;\;\;\mbox{(without
torsion)}
\end{equation}
In both cases the frequencies are real and  the strings
experience completely regular oscillatory behaviour.
Moreover, for small $n$ ($n<<mH\alpha'$) and large $n$ ($n>>mH\alpha'$),
the results agree, while there is a minor
difference in the intermediate region; in fact, from 
eqs.(\ref{corref})-(\ref{sintors}), [see also eqs.(5.35)-(5.36)]
follow that the effect of the conformal invariance
is to `complete the square'. This effect shows itself too in the mass 
spectrum [eqs.(\ref{especma}) and (\ref{espmsint})].

Notice that states with the same eigenvalue of the number-operator do
not necessarily have the same mass\cite{all2}. 
This is the case both for the low-mass
states and the high-mass states. In the low-mass spectrum, the effect is just
like a fine-structure effect, while in the high-mass spectrum, the states are
completely mixed up. This is very different from  Minkowski
spacetime  where states with the same eigenvalue of the number-operator
always have the same mass.

Interestingly enough, conformal invariance fixes the value of the
spacetime curvature to be
$$
R = - {69 \over 26 \; \alpha'} \; .
$$

The paper is organized as follows. In Section 2, we review the WZWN
construction for the group $SL(2,R)$.
We consider the two parametrizations corresponding to global 2+1 Anti de
Sitter spacetime and 2+1 Black
Hole Anti de Sitter spacetime, respectively. In both cases, we read off the
metric and torsion.

In Section 3, we solve the classical string equations of motion and
constraints in the
above mentioned backgrounds, for the special configuration describing an
oscillating circular
string. We compare with the analogue results obtained in the absence of
torsion, and then extract
and discuss the {\it precise} effect of the conformal invariance (expressed
via the parallelizing
torsion). 

In Section 4, we perform a semi-classical quantization of the
oscillating circular strings,
obtained in Section 3. In this way we obtain the semi-classical mass
spectrum. Again, we
compare with the analogue results obtained in the absence of torsion, and
extract
and discuss the {\it precise} effect of the conformal invariance. 

In Section 5, we consider more generic
string configurations by solving the classical string equations of motion
and constraints in a
perturbative scheme. We compute first and second order string-fluctuations
around the string center of
mass, and derive the classical mass formula. The frequencies of string
fluctuations are compared
with the analogue results obtained in the absence of torsion. 

In Section 6, we perform a canonical
quantization of the oscillator modes, and we derive the quantum mass
formula. The mass formula is
investigated in detail in different regimes, and we compare with the
results obtained using
semi-classical quantization in Section 4. In particular, we derive the
asymptotic level spacing and the
entropy of string states. 

Finally in Section 7, we give our concluding
remarks, and we discuss possible
continuations of our work.
\section{Classical Equations of Motion}
\setcounter{equation}{0}
To fix our notations and conventions, we give in this section
a short review of the WZWN construction for the group
$SL(2,R)$. This will lead to the classical string equations of motion in
the background of 2+1 dimensional Anti de Sitter (AdS) spacetime  with the
presence of parallelizing torsion. A different parametrization of the group
manifold leads
to the classical string equations of motion in
the background of 2+1 dimensional Black Hole Anti de Sitter (BH-AdS)
Spacetime  \cite{ban}
with the presence of parallelizing torsion.

Our starting point is the sigma-model action including the WZWN term at level
$k$ \cite{wit}: 
\begin{equation}
S_{\sigma}=-\frac{k}{4\pi}\int_{M} d\tau d\sigma\;\eta^{\alpha\beta}\mbox{Tr}[
g^{-1}\partial_\alpha g\;g^{-1}\partial_\beta g]-
\frac{k}{6\pi}\int_{B} \mbox{Tr}[
g^{-1}dg\wedge g^{-1}dg\wedge g^{-1}dg].
\end{equation}
Here $M$ is the boundary of the manifold $B$, and $g$ is a group-element of
$SL(2,R)$:
\begin{equation}
g=\left( \begin{array}{cc} a & u \\
-v & b \end{array}\right) ;\;\;\;\;\;\;\;\;\ ab+uv=1.
\end{equation}
Then, the action eq.(2.1) takes the form \cite{hor}:
\begin{equation}
S_{\sigma}=-\frac{k}{2\pi}\int_{M} d\tau d\sigma\;[\dot{a}\dot{b}-a'b'+
\dot{u}\dot{v}-u'v']-\frac{k}{\pi}\int_{M} d\tau d\sigma\;
\log (u) [\dot{a}b'-a'\dot{b}],
\end{equation}
where dot and prime denote derivative with respect to $\tau$ and $\sigma$,
respectively. We shall now consider a parametrization of the group manifold
corresponding to global 2+1 AdS. We first introduce new coordinates
$(X,Y,W,T)$:
\begin{equation}
a=H(W+X),\;\;\;\;\;\;b=H(W-X),\;\;\;\;\;\;u=H(T-Y),\;\;\;\;\;\;v=H(T+Y),
\end{equation}
where $H$ is a constant (the Hubble constant). Then we get from eq.(2.2):
\begin{equation}
X^2+Y^2-W^2-T^2=-\frac{1}{H^2},
\end{equation}
which is the standard embedding equation for 2+1 AdS.

Using the standard
parametrization (see for instance \cite{rind}):
\begin{eqnarray}
&X=r\cos\varphi,\;\;\;\;\;\;W=\frac{1}{H}\sqrt{1+H^2 r^2}\;\cos Ht,&\nonumber\\
&Y=r\sin\varphi,\;\;\;\;\;\;T=\frac{1}{H}\sqrt{1+H^2 r^2}\;\sin Ht,&
\end{eqnarray}
the action eq.(2.3)  becomes:
\begin{eqnarray}
S_{\sigma}&=&-\frac{kH^2}{2\pi}\int_{M} d\tau
d\sigma\;\left[ -(1+H^2r^2)(t'^2-\dot{t}^2)+\frac{r'^2-\dot{r}^2}{1+H^2r^2}+
r^2(\varphi'^2-\dot{\varphi}^2)\right] \nonumber\\
&-&\frac{kH^3}{\pi}\int_{M} d\tau
d\sigma\;r^2 \left[ \dot{t}\varphi'-t'\dot{\varphi}\right] .
\end{eqnarray}
Let us recall that the generic sigma-model action in the presence of metric
$g_{\mu\nu}$
and anti-symmetric
tensor $B_{\mu\nu}$ is:
\begin{equation}
S_{\sigma}=\frac{1}{2\pi\alpha'}\int_{M} d\tau
d\sigma\;\left[ g_{\mu\nu}(\dot{X}^\mu\dot{X}^\nu-X'^\mu X'^\nu)+
2B_{\mu\nu}(\dot{X}^\nu X'^\mu-\dot{X}^\mu X'^\nu)\right] .
\end{equation}
In our case, $X^\mu=(t,r,\varphi)$ and we can then read off:
\begin{eqnarray}
&g_{tt}=-(1+H^2r^2),\;\;\;\;\;\;g_{rr}=(1+H^2r^2)^{-1},\;\;\;\;\;\;
g_{\varphi\varphi}=r^2,&\nonumber\\
&B_{t\varphi}=-B_{\varphi t}=\frac{1}{2}Hr^2,
\end{eqnarray}
while the level of the WZWN model is related to the string tension 
and $ H $ through
\begin{equation}\label{k}
k=(H^2\alpha')^{-1}\; .
\end{equation}
Thus, the background is 2+1 AdS in static coordinates
(which cover AdS completely), plus an anti-symmetric tensor $B_{\mu\nu}$
with a single non-zero component $B_{t\varphi}$.

Alternatively we can parametrize the group-element eq.(2.2) in the
following way
\cite{kal}:
\begin{eqnarray}
&a=\frac{Hr}{\sqrt{M}}e^{\sqrt{M}\varphi},\;\;\;\;\;\;u=\pm\sqrt{\frac{H^2r^2-M}
{M}}\;e^{H\sqrt{M}t},&\nonumber\\
&b=\frac{Hr}{\sqrt{M}}e^{-\sqrt{M}\varphi},\;\;\;\;\;\;v=\pm\sqrt{\frac{H^2r
^2-M}
{M}}\;e^{-H\sqrt{M}t},&
\end{eqnarray}
where $M$ is a constant. This parametrization holds for $H^2r^2-M>0$, but
analogue expressions
hold for
$H^2r^2-M<0$.

With this parametrization, the action eq.(2.3) becomes:
\begin{eqnarray}
S_{\sigma}&=&-\frac{kH^2}{2\pi}\int_{M} d\tau
d\sigma\;\left[ -(H^2r^2-M)(t'^2-\dot{t}^2)+\frac{r'^2-\dot{r}^2}{H^2r^2-M}+
r^2(\varphi'^2-\dot{\varphi}^2)\right] \nonumber\\
&-&\frac{kH^3}{\pi}\int_{M} d\tau
d\sigma\;r^2 \left[ \dot{t}\varphi'-t'\dot{\varphi}\right] .
\end{eqnarray}
This is of course equivalent to the AdS-action (2.7) in the case $M=-1$.
However, for positive $M$, the background corresponding to eq.(2.12) is:
\begin{eqnarray}
&g_{tt}=-(H^2r^2-M),\;\;\;\;\;\;g_{rr}=(H^2r^2-M)^{-1},\;\;\;\;\;\;
g_{\varphi\varphi}=r^2,&\nonumber\\
&B_{t\varphi}=\frac{1}{2}Hr^2,\;\;\;\;\;\;B_{\varphi t}=-\frac{1}{2}Hr^2,
\end{eqnarray}
which is the 2+1 BH-AdS spacetime \cite{ban}
plus an anti-symmetric tensor $B_{\mu\nu}$
with a single non-zero component $B_{t\varphi}$.
And again the level of the WZWN model is $k=(H^2\alpha')^{-1}$ \cite{wel}.
We also recall that $M$ is
the mass of the black hole while $H$ is the Hubble constant.

We close this section with some general remarks concerning the action eq.(2.8).
The corresponding equations of motion  are:
\begin{equation}
\ddot{X}^\mu-X''^\mu+\Gamma^\mu_{\rho\sigma}(\dot{X}^\rho\dot{X}^\sigma-
X'^\rho X'^\sigma)+H^\mu_{\rho\sigma}(\dot{X}^\rho X'^\sigma-
\dot{X}^\sigma X'^\rho)=0,
\end{equation}
where, as usual, $H_{\mu\rho\sigma}=B_{\mu\rho,\sigma}-B_{\mu\sigma,\rho}+
B_{\rho\sigma,\mu}$.

The string equations of motion should be supplemented by the
constraints:
\begin{equation}
g_{\mu\nu}\dot{X}^\mu X'^\nu=0,\;\;\;\;\;\;g_{\mu\nu}(\dot{X}^\mu\dot{X}^\nu+
X'^\mu X'^\nu)=0.
\end{equation}
It is convenient to introduce world-sheet light-cone coordinates:
\begin{equation}
\sigma^\pm=\tau\pm\sigma.
\end{equation}
Then,  eq.(2.14) takes the compact form:
\begin{equation}
X^\lambda_{-}\bar{\nabla}_\lambda X^\mu_{+}=0,
\end{equation}
where $\bar{\nabla}_\lambda$ is the generalised covariant derivative
defined in terms of the generalised Christoffel symbol:
\begin{equation}
\bar{\Gamma}^\mu_{\rho\sigma}=\Gamma^\mu_{\rho\sigma}+H^\mu_{\rho\sigma}.
\end{equation}
Notice that $\bar{\Gamma}^\mu_{\rho\sigma}$ is obviously not symmetric in the
two lower indices.
\section{Circular Strings. Classical Dynamics}
\setcounter{equation}{0}
To investigate the effect of the conformal invariance
on the string dynamics, we shall first
consider the special string configurations representing oscillating
circular strings.

We consider the background eq.(2.9), corresponding to the 2+1
AdS spacetime; the results in the background eq.(2.13), corresponding to
the 2+1 BH-AdS spacetime, can
then be obtained immediately.

The Ansatz describing oscillating circular strings is:
\begin{equation}
t=t(\tau),\;\;\;\;\;\;r=r(\tau),\;\;\;\;\;\;\varphi=\sigma.
\end{equation}
Then, equations (2.14)-(2.15) lead to:
\begin{equation}
\ddot{t}+\frac{2H^2r\dot{r}\dot{t}}{1+H^2r^2}+\frac{2Hr\dot{r}}{1+H^2r^2}=0,
\end{equation}
\begin{equation}
\ddot{r}+(1+H^2r^2)H^2r\dot{t}^2+(1+H^2r^2)r-\frac{H^2r\dot{r}^2}{1+H^2r^2}+
2(1+H^2r^2)Hr\dot{t}=0,
\end{equation}
supplemented by the constraint:
\begin{equation}
-(1+H^2r^2)\dot{t}^2+\frac{\dot{r}^2}{1+H^2r^2}+r^2=0.
\end{equation}
These three equations (3.2)-(3.4) are consistently integrated to:
\begin{equation}
\dot{t}=\frac{E-Hr^2}{1+H^2r^2},
\end{equation}
\begin{equation}
\dot{r}^2=-(1+2EH)r^2+E^2,
\end{equation}
where $E$ is a non-negative integration constant. Equation (3.6) can be
conveniently written as:
\begin{equation}
\dot{r}^2+V(r)=0;\;\;\;\;\;\;\;\;V(r)=(1+2EH)r^2-E^2,
\end{equation}
that is, the potential $V(r)$ is {\it quadratic} in $r$. This is a great
simplification
as compared to the case of AdS without torsion. In that case \cite{all1},
the potential
was {\it quartic} in $r$ and given by:
\begin{equation}
V(r)=(1+H^2r^2)r^2-E^2,\;\;\;\;\;\;\;\;\;\;(\mbox{without torsion})
\end{equation}
that is, the solution involved elliptic functions \cite{all1}.
In the present case with conformal invariance, the solution is instead
obtained in terms of trigonometric
functions (see later). Thus, an effect of the conformal invariance is that
the mathematics simplifies
considerably. It is also interesting to notice that the torsion,
corresponding to conformal
invariance, gives rise to repulsion at large distances, while gravity
itself gives rise to
attraction in AdS. This follows from
a comparison of the  two potentials eqs.(3.7)-(3.8). It is seen from these
expressions that the
parallelizing torsion provides the term
$-H^2r^4$  for large $r$, i.e., a repulsive term in the potential. In fact,
this repulsive term
precisely cancels the dominant attractive term in the potential eq.(3.8) in
the absence of torsion.
The final
outcome,  in the presence of conformal invariance, is
that the potential, eq.(3.7), is still attractive, but it is only {\it
quadratic} in $r$.

As for the dynamics of the circular strings in the presence of
conformal invariance, we see from eq.(3.7) that for a given value of $E$
(and fixed $H$),
the string oscillates between $r=0$ and $r=r_{\mbox{max}}$:
\begin{equation}
r_{\mbox{max}}=\sqrt{\frac{E^2}{1+2EH}};\;\;\;\;\;\;E\geq 0
\end{equation}
Notice also that $\dot{t}$ is always positive during the oscillations.

In the case of circular strings in the background of 2+1 BH-AdS, eq.(2.13),
one finds
in a similar way:
\begin{equation}
\dot{t}=\frac{E-Hr^2}{H^2r^2-M},
\end{equation}
\begin{equation}
\dot{r}^2=(M-2EH)r^2+E^2.
\end{equation}
Then, the potential is:
\begin{equation}
\dot{r}^2+V(r)=0;\;\;\;\;\;\;V(r)=-(M-2EH)r^2-E^2,
\end{equation}
which is again {\it quadratic} in $r$.

In the 2+1 BH-AdS spacetime,
there is an event horizon at $r_{\mbox{hor}}=\sqrt{M}/H$, and we
demand that $\dot{t}\geq 0$ everywhere outside the horizon. This leads to the
constraint on the integration constant $E$:
\begin{equation}
E>\frac{M}{H}.
\end{equation}
It follows that for a given value of $E$ fulfilling eq.(3.13), a circular string
has a maximal radius  $r=r_{\mbox{max}}$:
\begin{equation}
r_{\mbox{max}}=\sqrt{\frac{E^2}{2EH-M}},
\end{equation}
it then contracts, crosses the horizon and falls into the black hole.
Qualitatively, this is the same behaviour as in the absence of torsion
\cite{all1}. But also in this case of BH-AdS, the conformal invariance
simplifies the
mathematics. More precisely, as in the case of AdS, the torsion
corresponding to conformal
invariance provides a repulsive term, which precisely cancels the dominant
attractive term obtained from gravity.
\section{Circular Strings. Semi-Classical Quantization}
\setcounter{equation}{0}
In this section we perform a semi-classical quantization of the circular
string configurations in the 2+1 AdS spacetime, obtained in the previous
section. We use an approach
developed in field theory by Dashen et. al. \cite{das} (see also
\cite{vega1}), based
on the stationary phase approximation of
the functional integral. In our context, this is supposed to be a good
approximation
in the "semi-classical" regime where $H^2\alpha'<<1$.

The method can be
only used for time-periodic solutions of
the classical equations of motion. Thus it can be used for the oscillating
circular strings in the 2+1 AdS spacetime. On the other hand, the circular
strings in the 2+1 BH-AdS are not truely time-periodic because of the
causal properties of the background: once the strings have passed the
horizon, they will not re-appear (although the solutions are formally
time-periodic from the mathematical point of view).

The result of the stationary phase integration is expressed in terms of
the function $W(m)$ \cite{das}:
\begin{equation}
W(m)\equiv S_{\mbox{cl}}(T(m))+m\;T(m),
\end{equation}
where $S_{\mbox{cl}}$ is the action of the classical solution, $m$ is the
mass and
the period $T(m)$ is implicitly given by:
\begin{equation}
\frac{dS_{\mbox{cl}}}{dT}=-m.
\label{extre}
\end{equation}
Here it is important that $T$ is the period in a {\it physical} time
variable. In our
case, it will be the period in the target-space time $t$.
The bound state quantization condition then becomes \cite{das}:
\begin{equation}
W(m)=2\pi n;\quad n \in N.
\label{concua}
\end{equation}
This condition is generally expected to hold for $n$ "large". In our case,
this will correspond
to (say) $n>>H^2\alpha'$.

We now use this method on the oscillating
circular strings in 2+1 AdS, as described by eqs.(3.5)-(3.6). These
equations are solved by:
\begin{equation}
r(\tau)=\frac{E}{\sqrt{1+2EH}}|\sin[\sqrt{1+2EH}\;\tau]|,
\end{equation}
\begin{equation}
Ht(\tau)=\mbox{arctan}
\left(\frac{1+EH}{\sqrt{1+2EH}}\tan[\sqrt{1+2EH}\;\tau]\right)
-\tau,
\end{equation}
where we took initial conditions such that:
\begin{equation}
t(0)=0,\;\;\;\;\;\;r(0)=0.
\end{equation}
The period of the solution, which is twice the period of $r$, is in world-sheet
time $\tau$ given by:
\begin{equation}
T_\tau=\frac{2\pi}{\sqrt{1+2EH}}.
\end{equation}
The classical action over one period is obtained from eq.(2.8), using
eq.(2.9) and
eqs.(4.4)-(4.5):
\begin{eqnarray}
S_{\mbox{cl}}&=&\frac{1}{2\pi\alpha'}\int_{0}^{2\pi} d\sigma\int_{0}^{T_\tau}
 d\tau\;\left[ g_{\mu\nu}(\dot{X}^\mu\dot{X}^\nu-X'^\mu X'^\nu)+
2B_{\mu\nu}(\dot{X}^\nu X'^\mu-\dot{X}^\mu X'^\nu)\right] \nonumber\\
&=&-\frac{2}{\alpha'}\int_{0}^{T_\tau}
 d\tau\;\frac{(1+EH)r^2}{1+H^2r^2}\nonumber\\
&=&-\frac{4\pi(1+EH)}{H^2\alpha'}\left[
\frac{1}{\sqrt{1+2EH}}-\frac{1}{1+EH}\right] .
\end{eqnarray}
As explained after eq.(4.2), the period $T$ must be the period in the
physical time $t$.
This period is obtained from eq.(4.7) and eq.(4.5):
\begin{equation}
T_{t}=\frac{2\pi}{H}\left(1-\frac{1}{\sqrt{1+2EH}}\right).
\end{equation}
Then:
\begin{equation}
S_{\mbox{cl}}(T_{t})=-\frac{1}{2\pi\alpha'}
T_{t}^2\; \left(1-\frac{HT_{t}}{2\pi}\right)^{-1} \; .
\end{equation}
From eq.(4.2) we can then obtain the mass:
\begin{equation}
m=-\frac{dS_{\mbox{cl}}}{dT}=\frac{1}{2\pi\alpha'}\left(2\,T_{t}-
\frac{HT_{t}^2}{2\pi}\right)
\left(1-\frac{HT_{t}}{2\pi}\right)^{-2},
\end{equation}
which can be inverted to obtain the physical period $T_{t}$:
\begin{equation}
T_{t}=2\pi\;\frac{\sqrt{1+Hm\alpha'}-1}{H\sqrt{1+Hm\alpha'}}.
\end{equation}
Finally, the quantization condition eqs.(4.1)-(4.3) becomes:
\begin{equation}
W(m)\equiv S_{\mbox{cl}}(T(m))+m\;T(m)=2\pi n,
\end{equation}
i.e.:
\begin{equation}
\frac{2\pi}{H^2\alpha'}\left(\sqrt{1+mH\alpha'}-1\right)^2=2\pi n.
\end{equation}
This equation can be solved for $m$ giving:
\begin{equation}
\alpha' m^2=4n\left(1+\frac{\sqrt{H^2\alpha' n}}{2}\;\right)^2,
\end{equation}
which gives the spectrum of quantum string states.

Notice that for "small" $n$
 ($n<<(H^2\alpha')^{-1}$) it gives:
\begin{equation}
\alpha' m^2=4n,
\end{equation}
which is the Minkowski result, while for "large" $n$ ($n>>(H^2\alpha')^{-1}$):
\begin{equation}
\alpha' m^2=H^2\alpha' n^2.
\end{equation}
These results must be compared with the analogue results obtained for
circular strings
 in AdS but {\it without} including
torsion \cite{vega2}. As a general effect, we see that in the presence
of conformal 
invariance, the mathematics is much simpler. However, the physics is
more or less {\it unchanged}. In fact, in the two limits
$n<<(H^2\alpha')^{-1}$ and $ n>>(H^2\alpha')^{-1}$, the results
obtained here are in {\it exact} agreement with the results obtained
without torsion \cite{vega2}. That is, for small $n$ and large $n$, the
spectrum is not affected by the conformal invariance, while there may be
some minor changes in the intermediate region.
\section{Perturbations Around String Center of Mass}
\setcounter{equation}{0}
The results in Chapters 3, 4 were obtained for  special string
configurations. In order to see if these results are more generic or
just particular 
to the circular strings, we must consider more general string configurations.

The equations of motion and constraints eqs.(2.14)-(2.15) can in principle
be solved exactly in the case of $SL(2,R)$, since it is a group-manifold
\cite{mic,eic}, but the formulas (see for instance \cite{bars}) are
formal and not explicit enough for further investigations of the
string dynamics. Instead, we shall use here the method of expansion around
the string center of mass \cite{san1}, that is, we will compute first and
second order string fluctuations around the point-particle geodesic
representing the center of mass of the string.
In the first subsection, we consider a generic 3-D spacetime
with arbitrary torsion. This subsection is thus the generalization of
subsection III.A
in Ref.\cite{all1} to the case of a spacetime with torsion. Then, in the
following subsection, we specialise to the case of 2+1 AdS with
parallelizing  torsion.
\subsection{General Formalism}
To be more precise, consider first the equations of motion eq.(2.14); the
constraints will be dealt with afterwards. We then expand \cite{san1}:
\begin{equation}
X^\mu(\tau,\sigma)=q^\mu(\tau)+\eta^\mu(\tau,\sigma)+\xi^\mu(\tau,\sigma)+...
\end{equation}
where $q^\mu(\tau)$ represents the string center of mass, while
$\eta^\mu(\tau,\sigma)$
and $\xi^\mu(\tau,\sigma)$ are the first and second order string
perturbations, respectively.

After insertion into  eq.(2.14), the equations of motion are to
be solved order by order in the expansion.
To zeroth order we get:
\begin{equation}
\dot{q}^\lambda\nabla_\lambda\dot{q}^\mu=0,
\end{equation}
which is just the standard general relativity geodesic equation; obviously the
torsion does not couple to the string center of mass. To first order in the
expansion, we get after a little algebra the following equation for
$\eta^\mu(\tau,\sigma)$:
\begin{equation}
\dot{q}^\lambda\bar{\nabla}_\lambda(\dot{q}^\delta\bar{\nabla}_\delta\eta^\mu)-
\bar{R}^\mu_{\sigma\rho\lambda}\dot{q}^\rho\dot{q}^\sigma\eta^\lambda-
\eta''^\mu=2H^\mu_{\rho\sigma}\dot{q}^\rho(\dot{q}^\delta\bar{\nabla}_\delta
\eta^\sigma-\eta'^\sigma),
\end{equation}
where $\bar{R}^\mu_{\sigma\rho\lambda}$ is the generalised curvature defined
via the generalised Christoffel symbol eq.(2.18):
\begin{equation}
\bar{R}^\lambda_{\mu\nu\beta}=\bar{\Gamma}^\lambda_{\mu\nu,\beta}-
\bar{\Gamma}^\lambda_{\mu\beta,\nu}+\bar{\Gamma}^\alpha_{\mu\nu}
\bar{\Gamma}^\lambda_{\alpha\beta}-\bar{\Gamma}^\alpha_{\mu\beta}
\bar{\Gamma}^\lambda_{\alpha\nu}.
\end{equation}
Notice that eq.(5.3) is a special case of the generalised Raychaudhury
equation for strings
in the presence of torsion \cite{kar}. Moreover,  if we
skip the
$\eta'$ and
$\eta''$ terms in eq.(5.3), then it is  just the generalised geodesic
deviation equation
(see for instance \cite{nash}).

However, we
can simplify eq.(5.3) further. For a massive string, corresponding to the
string center of mass fulfilling:
\begin{equation}
g_{\mu\nu}\dot{q}^\mu\dot{q}^\nu=-m^2\alpha'^2,
\end{equation}
there are two physical polarizations
of string perturbations around the geodesic $q^\mu(\tau)\;$ (we are in a
3-D spacetime). We therefore
introduce two normal vectors $n^\mu_R\;(R=1,2)$:
\begin{equation}
g_{\mu\nu}n^\mu_R\dot{q}^\nu=0,\;\;\;g_{\mu\nu}n^\mu_R n^\nu_S=\delta_{RS},
\end{equation}
and consider only first order perturbations in the form:
\begin{equation}
\eta^\mu= n^\mu_R \Phi^{R},
\end{equation}
where $\Phi^R$ are the comoving perturbations, i.e., the perturbations
as seen by an observer travelling with the center of mass of the string.
It must be noticed that for a string in a three dimensional spacetime,
there is only one
physical polarization
of string perturbations (one transverse direction), but since our zeroth
order solution is not a string but a point-particle, we get in some sense
one polarization of perturbations too many at this stage.
This extra degree of freedom will
eventually have
to be eliminated somehow using the constraints. Notice also that the
normal vectors eq.(5.6) are not uniquely defined. In fact, there
is a gauge invariance originating from the freedom to make local
rotations of the 2-bein spanned by the normal vectors. By generalising the
procedure of Ref.\cite{all1} to the case with torsion, we fix this gauge by
taking
the normal vectors to fulfil:
\begin{equation}
\dot{q}^\mu\bar{\nabla}_\mu n^\nu_R=0.
\end{equation}
Using eqs.(5.5)-(5.8) in eq.(5.3), we find after contraction with
$g_{\mu\nu}n^\nu_S$:
\begin{equation}
\ddot{\Phi}_S-\Phi''_S-\bar{R}_{\mu\sigma\rho\lambda}
n^\mu_S n^\lambda_R
\dot{q}^\rho\dot{q}^\sigma\Phi^R=2H_{\mu\rho\sigma}\dot{q}^\rho
n^\mu_S n^\sigma_R(\dot{\Phi}^R-\Phi'^R),
\end{equation}
which for a given background $(g_{\mu\nu}, B_{\mu\nu}$ must be solved for
$\Phi^R$.

For the
second order perturbations,
the picture is a little more complicated
since they couple also to the first order perturbations. We therefore
consider the full set
of perturbations $\xi^\mu$:
\begin{equation}
\dot{q}^\lambda\bar{\nabla}_\lambda(\dot{q}^\delta\bar{\nabla}_\delta\xi^\mu)-
\bar{R}^\mu_{\sigma\rho\lambda}\dot{q}^\rho\dot{q}^\sigma\xi^\lambda-
\xi''^\mu-2H^\mu_{\rho\sigma}\dot{q}^\rho(\dot{q}^\delta\bar{\nabla}_\delta
\xi^\sigma-\xi'^\sigma)=U^\mu.
\end{equation}
The term  $U^\mu$, which is bilinear in the first order perturbations,
plays the role of a source and is
explicitly given by:
\begin{eqnarray}
U^\mu&=&-\Gamma^\mu_{\rho\sigma}(\dot{\eta}^\rho\dot{\eta}^\sigma-
\eta'^\rho\eta'^\sigma)-2H^\mu_{\rho\sigma}\dot{\eta}^\rho \eta'^\sigma-
2\Gamma^\mu_{\rho\sigma,\lambda}\dot{q}^\rho
\eta^\lambda\dot{\eta}^\sigma\nonumber\\
&-&2H^\mu_{\rho\sigma,\lambda}\dot{q}^\rho
\eta^\lambda\eta'^\sigma-\frac{1}{2}\Gamma^\mu_{\rho\sigma,
\lambda\delta}\dot{q}^\rho\dot{q}^\sigma\eta^\lambda\eta^\delta.
\end{eqnarray}
After solving eqs.(5.9)-(5.10) for the first and second order
perturbations, the constraints eq.(2.15) have to be imposed. In world-sheet
light cone coordinates $\sigma^\pm=\tau\pm\sigma,$
the constraints take the form:
\begin{equation}
T_{\pm\pm}=g_{\mu\nu}\partial_\pm X^\mu\partial_\pm X^\nu=0.
\end{equation}
The world-sheet energy-momentum tensor $T_{\pm\pm}$ is conserved, as can be
easily verified using eq.(2.14), and therefore can be written:
\begin{equation}
T_{--}=\frac{1}{2\pi}\sum_n\tilde{L}_n e^{-in(\sigma-\tau)},\;\;\;T_{++}=
\frac{1}{2\pi}\sum_n L_n e^{-in(\sigma+\tau)}.
\end{equation}
At the classical level, the constraints are
then simply :
\begin{equation}
L_n=\tilde{L}_n=0;\;\;\; n\in Z.
\end{equation}
The quantum constraints will be considered in Section 6.
Up to second order in the expansion around the string center of mass we find:
\begin{eqnarray}
T_{\pm\pm}\hspace*{-2mm}&=&\hspace*{-2mm}-\frac{1}{4}m^2\alpha'^2+g_{\mu\nu}
\dot{q}^\mu\partial_\pm\eta^\nu
+\frac{1}{4}g_{\mu\nu,\rho}\dot{q}^\mu
\dot{q}^\nu\eta^\rho\nonumber\\
\hspace*{-2mm}&+&\hspace*{-2mm}g_{\mu\nu}\dot{q}^\mu\partial_\pm\xi^\nu
+g_{\mu\nu}\partial_\pm\eta^\mu\partial_\pm
\eta^\nu+g_{\mu\nu,\rho}\dot{q}^\mu\eta^\rho\partial_\pm\eta^\nu\nonumber\\
\hspace*{-2mm}&+&\hspace*{-2mm}\frac{1}{4}g_{\mu\nu,\rho}\dot{q}^\mu
\dot{q}^\nu\xi^\rho+\frac{1}{8}g_{\mu\nu,\rho\sigma}\dot{q}^\mu
\dot{q}^\nu\eta^\rho\eta^\sigma.
\end{eqnarray}
Formally, this is the same expression as in the absence of torsion, but one
should keep in mind that the solutions for $\eta$ and $\xi$ involve
the torsion and are different now.

Notice also that all results derived in this subsection hold for arbitrary
torsion
(not necessarily parallelizing).
In the next subsection we apply the above formalism to the case of strings
in the 2+1 AdS spacetime with  parallelizing torsion.
\subsection{Strings in 3-D AdS with  Parallelizing Torsion}
We now consider strings in the 2+1 AdS with  conformal invariance,
as described by
the metric and torsion eq.(2.9). For simplicity we consider a string with
radially
moving center of mass:
\begin{equation}
t=t(\tau),\;\;\;\;\;\;r=r(\tau),\;\;\;\;\;\;\varphi=\mbox{const.}
\end{equation}
Then, eqs.(5.2), (5.5) are integrated to:
\begin{equation}
\dot{t}=\frac{E}{1+H^2r^2},
\end{equation}
\begin{equation}
\dot{r}^2=E^2-(1+H^2r^2)m^2\alpha'^2,
\end{equation}
where $E$ is an integration constant (not the same as in eqs.(3.5)-(3.6)).
These
equations are solved in terms of trigonometric functions, but we shall not
need the explicit expressions here. A pair of independent normal-vectors
fulfilling eq.(5.6)
is provided by:
\begin{eqnarray}
N^\mu_\perp&=&\left(0,\;0,\;\frac{1}{r}\right),\nonumber\\
N^\mu_\parallel&=&\left(\frac{\dot{r}}{m\alpha'(1+H^2r^2)},\;
\frac{E}{m\alpha'},\;0\right).
\end{eqnarray}
However, they do not fulfil the gauge-condition eq.(5.8). We therefore make
a local
rotation and define normal-vectors $(n^\mu_1,n^\mu_2)$, fulfilling also
eq.(5.8), by:
\begin{equation}
\left( \begin{array}{c} n^\mu_1 \\ n^\mu_2  \end{array}\right)=
\left( \begin{array}{cc} \cos(mH\alpha'\tau) & -\sin(mH\alpha'\tau) \\
\sin(mH\alpha'\tau) & \cos(mH\alpha'\tau) \end{array}\right)
\left( \begin{array}{c} N^\mu_\perp \\ N^\mu_\parallel \end{array}\right).
\end{equation}
Moreover, for the background eq.(2.9):
\begin{equation}
\bar{R}^\lambda_{\mu\nu\beta}=0,
\end{equation}
which expresses the fact that the torsion is parallelizing for a group
manifold (see for
instance \cite{nash}). 

Then,  eq.(5.9) for the first order
perturbations reduces to:
\begin{equation}
\ddot{\Phi}_1-\Phi''_1+2mH\alpha'(\dot{\Phi}_2-\Phi_2')=0,
\end{equation}
\begin{equation}
\ddot{\Phi}_2-\Phi''_2-2mH\alpha'(\dot{\Phi}_1-\Phi_1')=0.
\end{equation}
Considering closed strings, we Fourier expand:
\begin{equation}
\Phi_R=\sum_n \phi_{Rn} e^{-in\sigma};\;\;\;\;\;\;\;R=1,2,
\end{equation}
so that eqs.(5.22)-(5.23) become:
\begin{equation}
\left( \begin{array}{c} \ddot{\phi}_{1n} \\ \ddot{\phi}_{2n} \end{array}
\right) + 2{\cal A}\left( \begin{array}{c} \dot{\phi}_{1n} \\
\dot{\phi}_{2n} \end{array}
\right) +{\cal B}\left( \begin{array}{c} {\phi}_{1n} \\
{\phi}_{2n} \end{array} \right)=
\left( \begin{array}{c} 0 \\ 0 \end{array}
\right),
\end{equation}
that is, two coupled ordinary linear differential equations of second order
with constant (matrix) coefficients ${\cal A},\;{\cal B}$:
\begin{equation}
{\cal A}=mH\alpha' \left( \begin{array}{cc} 0 & 1 \\ -1 & 0 \end{array}
\right),\;\;\;\;\;\;\;\;
{\cal B}=\left( \begin{array}{cc} n^2 & 2inmH\alpha' \\
-2inmH\alpha' & n^2 \end{array}
\right).
\end{equation}
The first order $\tau$-derivatives in eq.(5.25) are eliminated by a rotation
similar to eq.(5.20):
\begin{equation}
\left( \begin{array}{c} \phi_{1n} \\ \phi_{2n}  \end{array}\right)=
\left( \begin{array}{cc} \cos(mH\alpha'\tau) & -\sin(mH\alpha'\tau) \\
\sin(mH\alpha'\tau) & \cos(mH\alpha'\tau) \end{array}\right)
\left( \begin{array}{c} \hat{\phi}_{1n} \\ \hat{\phi}_{2n} \end{array}\right),
\end{equation}
such that:
\begin{equation}
\left( \begin{array}{c} \ddot{\hat{\phi}}_{1n}
\\ \ddot{\hat{\phi}}_{2n}  \end{array}\right)+
\left( \begin{array}{cc} n^2+m^2H^2\alpha'^2 & 2inmH\alpha' \\
-2inmH\alpha' & n^2+m^2H^2\alpha'^2 \end{array}\right)
\left( \begin{array}{c} \hat{\phi}_{1n} \\ \hat{\phi}_{2n} \end{array}\right)
=\left( \begin{array}{c} 0 \\ 0 \end{array}
\right),
\end{equation}
and the equations are then decoupled by a unitary transformation:
\begin{equation}
\left( \begin{array}{c} {\hat{\phi}}_{1n}
\\ {\hat{\phi}}_{2n}  \end{array}\right)=
U\left( \begin{array}{c} C_{1n}
\\ C_{2n}  \end{array}\right);\;\;\;\;\;\;\;\;U=\frac{1}{\sqrt{2}}
\left( \begin{array}{cc} i & -i \\
1 & 1 \end{array}\right).
\end{equation}
Then, we get:
\begin{equation}
\left( \begin{array}{c} \ddot{C}_{1n}
\\ \ddot{C}_{2n}  \end{array}\right)+
\left( \begin{array}{cc} (n+mH\alpha')^2 & 0 \\
0 & (n-mH\alpha')^2 \end{array}\right)
\left( \begin{array}{c} {C}_{1n} \\ {C}_{2n} \end{array}\right)
=\left( \begin{array}{c} 0 \\ 0 \end{array}
\right),
\end{equation}
which are solved by:
\begin{eqnarray}
&C_{1n}=A_{1n}e^{-i|n+mH\alpha'|\tau}+
\tilde{A}_{1n}e^{i|n+mH\alpha'|\tau},&\nonumber\\
&C_{2n}=A_{2n}e^{-i|n-mH\alpha'|\tau}+
\tilde{A}_{2n}e^{i|n-mH\alpha'|\tau},&
\end{eqnarray}
where $(A_{Rn},\tilde{A}_{Rn})$ are integration constants.

The final result for the first order comoving perturbations is then:
\begin{equation}
\left( \begin{array}{c} \Phi_1
\\ \Phi_2  \end{array}\right)=\frac{1}{\sqrt{2}}\sum_n e^{-in\sigma}
\left( \begin{array}{cc} ie^{imH\alpha'\tau} & -ie^{-imH\alpha'\tau} \\
e^{imH\alpha'\tau} & e^{-imH\alpha'\tau} \end{array}\right)
\left( \begin{array}{c} C_{1n}
\\ C_{2n} \end{array}\right) .
\end{equation}
The constants $(A_{Rn},\tilde{A}_{Rn})$ are constrained by the condition that
$(\Phi_1,\Phi_2)$ are real. This leads to:
\begin{equation}
\tilde{A}_{2n}=(A_{1-n})^\dagger,\;\;\;\;\;\;\;\;
A_{2n}=(\tilde{A}_{1-n})^\dagger.
\end{equation}
As for the first order perturbations $\eta^\mu$, we get:
\begin{equation}
\eta^\mu=\frac{1}{\sqrt{2}}\sum_n e^{-in\sigma}[
(N^\mu_\parallel+iN^\mu_\perp)C_{1n}+
(N^\mu_\parallel-iN^\mu_\perp)C_{2n}],
\end{equation}
in terms of the normal-vectors eq.(5.19) and
the oscillators eq.(5.31). This concludes the derivation
of the first order perturbations. Notice that the frequencies are shifted
away from integers $n$:
\begin{equation}
\omega_n=|n\pm mH\alpha'|,
\end{equation}
and that the frequencies are different in the two directions perpendicular
to the geodesic of the string center of mass.
It is interesting to compare with the similar result in 2+1 AdS but
without torsion \cite{all2}. In that case, the frequencies of the first order
perturbations turned out to be \cite{all2}:
\begin{equation}
\omega_n=\sqrt{n^2 + m^2H^2\alpha'^2},\;\;\;\;\;\;\;\;\;\;\mbox{(without
torsion)}
\end{equation}
Thus, in both cases the frequencies are real, and therefore the strings
experience completely regular oscillatory behaviour.
Moreover, for small $n$ ($n<<mH\alpha'$) and large $n$ ($n>>mH\alpha'$),
the results
agree, while there is a minor
difference in the intermediate region; in fact, from eqs.(5.35)-(5.36)
follow that the effect of the conformal invariance
is to "complete the square".

We now come to the second order perturbations $\xi^\mu$, as determined by
eqs.(5.10)-(5.11). The computations are now going to be somewhat more
complicated so we merely give the results of the different steps. We first
re-define the $\xi$'s and the corresponding sources $U$:
\begin{equation}
\xi^t=\hat{\xi}^t,\;\;\;\;\;\;\xi^r=(1+H^2r^2)\hat{\xi}^r,\;\;\;\;\;\;
\xi^\phi=\frac{1}{r}\hat{\xi}^\varphi,
\end{equation}
\begin{equation}
U^t=\hat{U}^t,\;\;\;\;\;\;U^r=(1+H^2r^2)\hat{U}^r,\;\;\;\;\;\;
U^\phi=\frac{1}{r}\hat{U}^\varphi.
\end{equation}
Equation (5.10) then takes the form:
\begin{equation}
\left( \begin{array}{c} \ddot{\hat{\xi}}^t \\ \ddot{\hat{\xi}}^r \\
\ddot{\hat{\xi}}^\varphi \end{array}
\right) - \left( \begin{array}{c} \hat{\xi}''^{t} \\ \hat{\xi}''^{r} \\
\hat{\xi}''^\varphi \end{array}
\right) + 2{\cal D}\left( \begin{array}{c} \dot{\hat{\xi}}^t \\
\dot{\hat{\xi}}^r \\ \dot{\hat{\xi}}^\varphi \end{array}
\right) +2{\cal E}\left( \begin{array}{c} \hat{\xi}'^{t} \\
\hat{\xi}'^{r} \\ \hat{\xi}'^\varphi \end{array}
\right) +{\cal F}\left( \begin{array}{c} {\hat{\xi}}^t \\
{\hat{\xi}}^r \\ {\hat{\xi}}^\varphi \end{array} \right)=
\left( \begin{array}{c} {\hat{U}}^t \\ {\hat{U}}^r \\
{\hat{U}}^\varphi \end{array}
\right),
\end{equation}
where the matrices ${\cal D}$, ${\cal E}$ and ${\cal F}$ are given by:
\begin{equation}
{\cal D}=\frac{H^2r}{1+H^2r^2} \left( \begin{array}{ccc}
\dot{r} & E & 0 \\ E & \dot{r} & 0 \\ 0 & 0 & 0 \end{array}
\right),
\end{equation}
\begin{equation}
{\cal E}=\frac{H}{1+H^2r^2} \left( \begin{array}{ccc}
0 & 0 & \dot{r} \\ 0 & 0 & E \\ (1+H^2r^2)\dot{r} & -(1+H^2r^2)E & 0 \end{array}
\right),
\end{equation}
\begin{equation}
{\cal F}=\frac{H^2}{1+H^2r^2} \left( \begin{array}{ccc}
0 & 2E\dot{r} & 0 \\ 0 & 2E^2-m^2\alpha'^2+H^2m^2 r^2 & 0
\\ 0 & 0 & (1+H^2r^2)m^2\alpha'^2  \end{array}
\right).
\end{equation}
The first order $\tau$-derivatives in eq.(5.39)
are eliminated by the transformation:
\begin{equation}
\left( \begin{array}{c} {\hat{\xi}}^t \\
{\hat{\xi}}^r \\ {\hat{\xi}}^\varphi \end{array} \right)={\cal G}
\left( \begin{array}{c} {\hat{\Sigma}}^t \\
{\hat{\Sigma}}^r \\ {\hat{\Sigma}}^\varphi \end{array} \right);\;\;\;\;\;\;
{\cal G}=\mbox{Exp}\left( -\int^\tau{\cal D}(\tau')\; d\tau'\right),
\end{equation}
that is:
\begin{equation}
{\cal G}=\frac{-1}{(1+H^2r^2)m\alpha'}\left( \begin{array}{ccc}
\dot{r} & E & 0 \\ E & \dot{r} & 0 \\ 0 & 0  & -(1+H^2r^2)m\alpha' \end{array}
\right).
\end{equation}
We now Fourier expand the second order perturbations and the sources:
\begin{equation}
\hat{\Sigma}^\mu(\tau,\sigma)=\sum_n\hat{\Sigma}^\mu_n(\tau)
e^{-in\sigma},
\end{equation}
\begin{equation}
\hat{U}^\mu(\tau,\sigma)=\sum_n \hat{U}^\mu_n(\tau)e^{-in\sigma}.
\end{equation}
Then, the matrix equation (5.39) reduces to:
\begin{equation}
\left( \begin{array}{c} \ddot{\hat{\Sigma}}^t_n \\
\ddot{\hat{\Sigma}}^r_n \\ \ddot{\hat{\Sigma}}^\varphi_n
\end{array}\right)+{\cal V}
\left( \begin{array}{c} \hat{\Sigma}^t_n \\ \hat{\Sigma}^r_n \\
\hat{\Sigma}^\varphi_n
\end{array}\right)={\cal G}^{-1}
\left( \begin{array}{c} \hat{U}^t_n \\ \hat{U}^r_n \\
\hat{U}^\varphi_n
\end{array}\right),
\end{equation}
where:
\begin{eqnarray}
{\cal V}&=&{\cal G}^{-1}\left( n^2 I+{\cal F}-{\cal D}^2-\dot{{\cal D}}-
2in{\cal E}\right) {\cal G}\nonumber\\
&=&\left( \begin{array}{ccc} n^2+m^2H^2\alpha'^2 & 0 & 2inmH\alpha' \\
0 & n^2 & 0 \\
-2inmH\alpha' & 0 & n^2+m^2H^2\alpha'^2
\end{array} \right).
\end{eqnarray}
Thus, the second order perturbations are determined by a set of three
coupled linear ordinary differential equations of second order with
constant (matrix) coefficients and a complicated source-term.
It follows that the complete solution is known explicitly:
The matrix, eq.(5.48), is diagonalised in the same way as in eq.(5.29).
The full solutions for the three
second order perturbations are then written as free wave parts with
frequencies $|n+mH\alpha'|,\;|n-mH\alpha'|$
and $n$, respectively, plus particular solutions involving integrals of the
sources. This concludes the derivation of the second order perturbations.

Having
calculated the first and second order perturbations, we can now also
calculate the world-sheet energy-momentum tensor $T_{\pm\pm}$,
eqs.(5.12)-(5.15).
This
calculation is simplified using the fact that $T_{\pm\pm}$ are functions of
$n(\sigma\pm\tau)$ while the first order perturbations $\eta^\mu$ are
functions of $(n\sigma\pm|n\pm\ mH\alpha'|\tau)$.
The first order perturbations can
therefore only give constant contributions to $T_{\pm\pm}$. It is then
straightforward to compute $L_0$ and $\tilde{L}_0$:
\begin{equation}
L_0=\pi\sum_n \left[ (|n+mH\alpha'|+n)^2 A_{1n}A_{1n}^\dagger+
(|n-mH\alpha'|+n)^2 A_{2n}A_{2n}^\dagger\right] -\frac{\pi}{2}m^2\alpha'^2,
\end{equation}
\begin{equation}
\tilde{L}_0=\pi\sum_n \left[ (|n+mH\alpha'|-n)^2 A_{1n}A_{1n}^\dagger+
(|n-mH\alpha'|-n)^2 A_{2n}A_{2n}^\dagger\right] -\frac{\pi}{2}m^2\alpha'^2.
\end{equation}
The constraints eq.(5.14) for $n=0$ then become:
\begin{equation}
\sum_n n \left[ |n+mH\alpha'|A_{1n}A_{1n}^\dagger+
|n-mH\alpha'|A_{2n}A_{2n}^\dagger\right] =0,
\end{equation}
as well as:
\begin{equation}
m^2\alpha'^2=2\sum_n \left[\left( (n+mH\alpha')^2+n^2\right)
A_{1n}A_{1n}^\dagger+
\left( (n-mH\alpha')^2+n^2\right) A_{2n}A_{2n}^\dagger\right] ,
\end{equation}
determining the mass of the string. Notice that the mass formula of the
string is modified with respect to the usual flat spacetime expression
$(m^2\alpha'^2=4\sum_n n^2[A_{1n}A_{1n}^\dagger+A_{2n}A_{2n}^\dagger]).$
The reason for this modification is of course the presence of the
cosmological constant through both gravity and torsion.
\section{The Quantum Mass-Formula}
\setcounter{equation}{0}
In this section we perform the canonical quantization using
the results of the previous section.
The first order comoving perturbations are described by the action
(compare with eqs.(5.22)-(5.23)):
\begin{equation}
S^{(2)}=-\frac{1}{4\pi\alpha'}\int d\tau d\sigma\;\eta^{ab}\delta_{RS}
\left(\Phi^R_{,a}+A_{aU}^{R}\Phi^U\right)
\left(\Phi^S_{,b}+A_{bV}^S\Phi^V\right),
\end{equation}
where the vector-potential $A_{a}^{RS}$ is anti-symmetric in the
$RS$-indices, and explicitly given by:
\begin{equation}
A_\tau^{12}=A_\sigma^{12}=mH.
\end{equation}
Again, it is interesting to compare with the analogue action in the absence
of torsion \cite{all2}.
In that case, the action for the comoving first order perturbations
involved a {\it scalar} potential.
Thus we see that the effect of the conformal invariance precisely is to
cancel this scalar potential and
replace it by a {\it vector} potential. This actually follows more
generally from eq.(5.9). The scalar
potential comes from the $\bar{R}_{\mu\sigma\rho\lambda}$-term, while the
vector potential comes from
the term on the right hand side. Then, in the absence of torsion in AdS,
the scalar potential survives
but there is no vector potential. On the other hand, with torsion
corresponding to conformal invariance
in AdS, the vector potential survives, but there is no scalar potential
since the torsion is
parallelizing ($\bar{R}_{\mu\sigma\rho\lambda}=0$).

The momentum conjugate
to
$\Phi^R$ is:
\begin{equation}
\Pi_R\equiv\frac{\delta S^{(2)}}{\delta(\dot{\Phi}^R)}=\frac{1}{2\pi
\alpha'}(\dot{\Phi}_R+mH\epsilon_{RS}\Phi^S);\;\;\;\;\;\;\;\;\epsilon_{12}=1.
\end{equation}
We now go directly to the quantum theory.
The canonical commutation relations become:
\begin{eqnarray}
&[\Phi^R,\;\Phi^S]=[\Pi_R,\;\Pi_S]=0,&\nonumber
\end{eqnarray}
\begin{eqnarray}
&[\Pi_R,\;\Phi^S]=-i\delta^S_R\delta(\sigma-\sigma').&
\end{eqnarray}
It follows that
the constants $A_{Rn}$ and $\tilde{A}_{R_n}$
introduced in eqs.(5.31)-(5.33),
which are now considered as quantum operators,
have the following commutation relations:
\begin{equation}
[A_{1n},\;A_{1n}^\dagger]=\frac{\alpha'}{2|n+mH\alpha'|},\;\;\;\;\;\;\;\;
[A_{2n},\;A_{2n}^\dagger]=\frac{\alpha'}{2|n-mH\alpha'|}.
\end{equation}
It is  convenient to make the redefinitions:
\begin{eqnarray}
&A_{1n}\equiv\left\{ \begin{array}{cl}
\tilde{a}^1_n \sqrt{\frac{\alpha'}{2|n+mH\alpha'|}} & n>0, \\
a^2_{-n} \sqrt{\frac{\alpha'}{2|n+mH\alpha'|}}
& n<0, \end{array}\right.&\nonumber
\end{eqnarray}
\begin{eqnarray}
&A_{2n}\equiv\left\{ \begin{array}{cl}
\tilde{a}^2_n \sqrt{\frac{\alpha'}{2|n-mH\alpha'|}} & n>0, \\
a^1_{-n} \sqrt{\frac{\alpha'}{2|n-mH\alpha'|}} & n<0, \end{array}\right.&
\end{eqnarray}
and similarly for the hermitean conjugates. The $a^R_n$ and $\tilde{a}^R_n$
represent conventionally normalized oscillators (no summation over $R$):
\begin{eqnarray}
&[a^R_n,\;(a^R_n)^\dagger]=[\tilde{a}^R_n,\;(\tilde{a}^R_n)^\dagger]=1,
&\;\;\;\;\;\;\mbox{for all}\;\;n>0\nonumber\\
&[A^R_0,\;(A^R_0)^\dagger]=\frac
{1}{2mH}.
\end{eqnarray}
The classical constraints $L_0=\tilde{L}_0=0$ in the quantum theory take
the form:
\begin{equation}
(L_0-2\pi\alpha' a)\mid\psi>=(\tilde{L}_0-2\pi\alpha' a)\mid\psi>=0,
\end{equation}
where $a$ is the normal-ordering constant and the factor $2\pi\alpha'$ is
introduced for later convenience. The normal-ordering constant is most
easily obtained by symmetrization of the oscillator products in
eqs.(5.49)-(5.50).

The physical state conditions eq.(6.8), in terms of the conventionally
normalized oscillators, then become:
\begin{eqnarray}\label{especma}
m^2\alpha'&=&\sum_{n>0}\left[\frac{(n+mH\alpha')^2+n^2}{|n+mH\alpha'|}
+\frac{(n-mH\alpha')^2+n^2}{|n-mH\alpha'|}\right]\nonumber\\
&+&\sum_{n>0}\left[\frac{(n+mH\alpha')^2+n^2}{|n+mH\alpha'|}
\left( (a^1_n)^\dagger a^1_n+(\tilde{a}^1_n)^\dagger \tilde{a}^1_n\right)
\right] \nonumber\\
&+&\sum_{n>0}\left[\frac{(n-mH\alpha')^2+n^2}{|n-mH\alpha'|}
\left( (a^2_n)^\dagger a^2_n+
(\tilde{a}^2_n)^\dagger \tilde{a}^2_n\right)\right],
\end{eqnarray}
and:
\begin{equation}
\sum_{n>0} n \left[ (a^R_n)^\dagger
a^R_n-(\tilde{a}^R_n)^\dagger\tilde{a}^R_n\right]=0,
\end{equation}
where the zero-modes have been eliminated.

Eq. (6.10) simply expresses that there must be an equal amount of
left-movers and right-movers, so let us now consider the quantum mass-formula
eq.(6.9) in a little more detail. The first term in eq.(6.9)
represents the zero-point energy. At the present stage it is formally
infinite and need to be renormalized, but since it is just
an overall constant,
we skip it for the moment and concentrate on the oscillator-parts of
eq.(6.9). As in Minkowski
spacetime, it is convenient to characterize the physical states by the
eigenvalue of the number-operator:
\begin{equation}
N=\frac{1}{2}\sum_{n>0} \left[ (a^R_n)^\dagger
a^R_n+(\tilde{a}^R_n)^\dagger\tilde{a}^R_n\right].
\end{equation}
Returning to eq.(6.9), we first notice that we get the correct
Minkowski spacetime result (as we should) in the limit $H=0$. Moreover,
for the low-mass states ($mH\alpha'<<1$), the spectrum is just the Minkowski
spectrum with small corrections of order $H^2\alpha'$ (we always assume
$H^2\alpha'<<1$):
\begin{equation}
m^2\alpha'=4N+{\cal O}(H^2\alpha');\;\;\;\;\;\;\;\;\mbox{(low-mass states)}
\end{equation}
and we skipped the zero-point energy. 

Consider now the high-mass states
($mH\alpha'>>1$). As an example, we consider the state:
\begin{equation}
\left[ (\tilde{a}^1_1)^\dagger(a^1_1)^\dagger\right]^{N} |0>,
\end{equation}
for some large $N$ (say) $N>>(H^2\alpha')^{-1}$.
This is a state with eigenvalue $N$ of the number-operator, and its
mass is approximately:
\begin{equation}
m^2\alpha'\approx 4H^2\alpha'N^2.
\end{equation}
As another example of a high-mass state, consider (for $N$ even):
\begin{equation}
\left[ (\tilde{a}^1_2)^\dagger(a^1_2)^\dagger\right]^{N/2} |0>.
\end{equation}
This state also has eigenvalue $N$ of the number-operator, but its
mass is approximately:
\begin{equation}
m^2\alpha'\approx H^2\alpha'N^2.
\end{equation}
More generally, we find for the high-mass states (up to a numerical factor):
\begin{equation}
m^2\alpha'\sim H^2\alpha'N^2;\;\;\;\;\;\;\;\;\mbox{(high-mass states)}
\end{equation}
Notice that states with the same eigenvalue of the number-operator
(for instance the states eqs.(6.13), (6.15)), do
not necessarily have the same mass. This is the case both for the low-mass
states and the high-mass states. In the low-mass spectrum, the effect is just
like a fine-structure, while in the high-mass spectrum, the states are
completely mixed up. This is very different from the case of Minkowski
spacetime  where states with the same eigenvalue of the number-operator
always have the same mass.

Finally, we should compare also with the results obtained for AdS but without
torsion \cite{all2}. In that case, the mass-formula was found to be:
\begin{equation}\label{espmsint}
m^2\alpha'=2\sum_{n>0}\frac{2n^2+m^2H^2\alpha'^2}{\sqrt{n^2+m^2H^2\alpha'^2}}
+\sum_{n>0}\frac{2n^2+m^2H^2\alpha'^2}{\sqrt{n^2+m^2H^2\alpha'^2}}
\left[ (a^R_n)^\dagger
a^R_n+(\tilde{a}^R_n)^\dagger\tilde{a}^R_n\right].
\end{equation}
It follows that the results agree in the low-mass spectrum and in the
high-mass spectrum, while there is a minor difference in the intermediate
region. In fact, as already discussed in connection with the frequencies of
the first order perturbations (see the discussion after eq.(5.36)),
the effect of the conformal invariance is to "complete the squares".

It means that the conclusions obtained in ref.\cite{all2}
for AdS without torsion generally
hold as well in the presence of torsion, corresponding to conformal invariance
(parallelizing torsion).
In particular, the scale of the high-mass states is set by the Hubble constant
$H$ (and not by $\alpha'$) as follows from eq.(6.17). 

Moreover, the level spacing corresponding to eq.(6.17),
grows proportionally to $N$:
\begin{equation}
\frac{d(m^2\alpha')}{dN}\propto N
\end{equation}
This is  contrary to the case of
Minkowski spacetime where it is constant. As shown in \cite{all2}, this implies
that the density of levels, $ \rho(m) $ grows like $ \sim e^{\sqrt{m/H}} $
and that the partition function for a gas of strings in AdS is well-defined 
for any temperature $ \beta^{-1} $ (with or without torsion).
That is, there is no Hagedorn temperature in AdS. 

It also follows from eq.(6.17) that the entropy is proportional to $\sqrt{m}$:
\begin{equation}
S\sim\sqrt{N}\sim\sqrt{m},
\end{equation}
where we skipped numerical factors. It is interesting to notice that this
result is formally similar to the
recent results of the entropy obtained for the quantization of the $2+1$
BH-AdS spacetime \cite{stromdan}.

Notice also that the results of the canonical quantization obtained in
this section, agree with the results obtained by semi-classical quantization
of circular strings obtained in Section 4.
\section{Conclusions}
\setcounter{equation}{0}
In conclusion, we have  considered classical and quantum strings in 2+1
dimensional Anti de Sitter spacetime with parallelizing  torsion, 
corresponding to the conformally invariant $SL(2,R)$-WZWN  background.

By considering special and generic string configurations classically and
quantum mechanically, we have extracted the {\it precise} effects of the 
conformal invariance, both on the classical dynamics of strings and on the 
quantum mass spectrum.

Generally, we have seen that the conformal invariance leads to a number of
mathematical simplifications. On the other hand, the physical properties turn 
out to be more or less unchanged, as compared to the case without torsion.

This means that the original results obtained in AdS spacetime without
torsion \cite{all1,vega2,all2} still hold in the presence of conformal
invariance, that is, in the presence of
parallelizing torsion. At the quantum level, this means in particular that
the  high-mass states are
governed by:
\begin{equation}
 m\sim HN;\;\;\;\;\;\;\;\; N\in N_{0},\;\;\;\;(N \;\;"\mbox{large}"),
\end{equation}
where $m$ is the string mass and $H$ is
the Hubble constant. It follows that the level spacing grows proportionally
to $N$:
\begin{equation}
\frac{d(m^2\alpha')}{dN}\sim N,
\end{equation}
 while the entropy goes like:
\begin{equation}
S\sim\sqrt{m}.
\end{equation}
 Moreover, it follows that there is no Hagedorn temperature, so that the
partition
function is well-defined at any temperature.

These results were obtained using two independent methods, namely
semi-classical quantization of
circular oscillating strings and canonical quantization of string oscillator modes. The results of the
two approaches agree and they agree with the results obtained for vanishing
torsion \cite{all1,vega2,all2}.

\bigskip

The central charge in the $2+1$AdS WZWN model  takes the 
value\cite{lin,nem,gko}:
$$
c = {3k \over k-2} \; .
$$
Conformal invariance thus holds for $ k $ such that $ c = 26 $. This leads
to the value for $k$:
\begin{equation}\label{kmame}
k = {52 \over 23} \; .
\end{equation}
This means that conformal invariance holds provided the string tension and 
$ H $ are related  as follows
$$
H = {1 \over \sqrt{k \; \alpha'}} = \sqrt{23 \over 52 \; \alpha'} \; .
$$
where we used eqs.(\ref{k}) and (\ref{kmame}).

Moreover, the scalar curvature takes then the  value,
$$
R = - {6 \over k \; \alpha'} = - {69 \over 26 \; \alpha'}\; .
$$

\vskip 12pt
It would be interesting to generalize our analysis to higher dimensional
Anti de Sitter spacetime
AdS$_{D}$. Contrary to the three dimensional case, AdS$_{D}$ can generally
not be described as a
group-manifold. It can however be represented as a coset-space:
\begin{equation}\label{adscos}
AdS_{D}=\frac{SO(D-1,2)}{SO(D-1,1)}.
\end{equation}
It follows that AdS$_{D}$ is generally not conformally invariant. However,
as shown in Ref.\cite{lin,nem},
conformal invariance can be achieved for certain values of the cosmological
constant. More precisely, the central charge of the WZWN model of level $k$
for the coset represented in eq.(\ref{adscos}), is \cite{lin,nem,gko}:
\begin{equation}\label{cdd}
C=\frac{k\,  D(D+1)}{2[k +1-D]}-\frac{k\,  D(D-1)}{2[k+2-D]},
\end{equation}
As shown in   Ref.\cite{lin}, the condition of conformal invariance,
\begin{equation}\label{c26}
C=26\; ,
\end{equation}
has solutions in an {\it arbitrary} number of dimensions. However, in each
dimension, the cosmological
constant must take very specific values. That is, eqs.(\ref{cdd}-\ref{c26}) 
lead to an equation of the form: $k=k(D)$ with solutions $D, k(D)$ for {\it
arbitrary} values of $D$.

For our purposes, we would have to consider the gauged WZWN models
corresponding to the coset-spaces
eq.(\ref{adscos}) \cite{bars,sfet2}. This would allow us to read off the
corresponding metric and
antisymmetric tensor, which are necessary for the investigation of string
dynamics in these backgrounds.
This is currently under investigation.

\end{document}